\documentclass[doublespacing]{elsart}

\usepackage{graphicx}
\usepackage{amssymb}

\begin{document}
\begin{frontmatter}
\journal{SCES'2001}

\title { Anisotropic Superconductivity in the Induced Pairing Model}

%
%
%
%
%

\author{R. Micnas,}{\corauthref{1}}
\author{S. Robaszkiewicz}
\author{and B.Tobijaszewska}

\corauth[1]{Corresponding Author: 
         Roman Micnas 
	 Institute of Physics,
	 A. Mickiewicz  University,
         Umultowska 85, 
	 PL-61614 Poznan, Poland;
	 Phone: (+48 61) 8273041, 
         Email: rom@alpha.amu.edu.pl}
%
 
\address{Institute of Physics,  A. Mickiewicz  University,
             Umultowska 85, 61-614 Pozna\'{n}, Poland }

%
%
%
%


\begin{abstract} 
The model of local electron pairs and itinerant fermions  coupled 
via charge exchange mechanism, which mutually induces superconductivity 
in both subsystems is studied for anisotropic pairing symmetry. 
The phase diagram  is presented and the phase fluctuations effects 
are analyzed within the Kosterlitz-Thouless scenario.

\end{abstract}

%
%



\end{frontmatter}

%
%
%
%
%
%
%

A mixture of interacting charged bosons 
  (bound electron pairs)  and  electrons  can  show  features  which  are 
  intermediate between those of  local pair (LP) superconductors and those 
  of classical BCS systems. Such a  two component model is of relevance
  for high temperature superconductors 
  \cite{{Micnas90},{gorkov87},{jrmr96},{larkin97}}.\\
  In this paper we  study  a generalization of the model to the case of
anisotropic pairing of extended s--wave or d--wave type, which 
is defined by the following  Hamiltonian 
\begin{eqnarray}\label{zero}
{\it H} = \sum_{\bf {k}\sigma} (\epsilon_{\bf k}-
\mu)c^{\dagger}_{\bf{k}\sigma}
c_{\bf {k}\sigma} + 2\sum_{i}(\Delta_{0}-\mu)b^{\dagger}_{i}b_{i} \nonumber\\
-\sum_{ij}J_{ij}b^{\dagger}_{i}b_{j} +\frac{1}{\sqrt{N}}\sum_{\bf q}I(B^{\dagger}_{\bf q}b_{\bf q} 
+b^{+}_{\bf q}B_{\bf q}),
\end{eqnarray}
$\epsilon_{\bf k}$  refers to  the  energy band  of  the  c-electrons, 
  $\Delta_{0}$   measures  the 
  relative position of the LP level with respect to the bottom of 
  the c-electron band. 
  $\mu$ is the chemical potential which ensures that the 
  total number of particles in the system is constant, i.e. 
$ n=\frac{1}{N}(\sum_{\bf k\sigma}
<c^{\dagger}_{\bf k \sigma}c_{\bf k \sigma}>
+2\sum_{i} <b^{\dagger}_{i}b_{i}>)=n_{c}+2n_{b}$.
$J_{ij}$ is the pair hopping integral.    
$B^{\dagger}_{\bf q}=\sum_{\bf k}\phi_{k}c^{\dagger}_{\bf{k} +\bf {q}/2,
\uparrow}c^{\dagger}_{\bf {-k} +\bf {q}/2 \downarrow}$ denotes 
 the singlet pair creation
operator of $c$-electrons and $I$ is the coupling constant.
 The  operators for local pairs (hard-core charged bosons)
$ {b^{\dagger}_{i},b_{i}}$  
obey the Pauli commutation rules: $[b_{i},b^{\dagger}_{j}]=
(1-2n_{i})\delta_{ij}, (b^{\dagger}_{i})^2=(b_{i})^2=0, b^{\dagger}_{i}b_{i}+
b_{i}b^{\dagger}_{i}=1, n_{i}=b^{\dagger}_{i}b_{i} $.
We assume that the coupling between the two subsystems is 
via the center of mass momenta ${\bf q}$ of the Cooper pair $B^{\dagger}_{\bf q}$
 and the hard-core boson $b_{\bf q}$. 
 The pairing symmetry, on a 2D square lattice,  is determined by the form of $\phi_{k}$, which is 1 for the on-site pairing,
 $\phi_{k}=\gamma_{k}=\cos(k_{x}a)+\cos(k_{y}a)$ for the extended $s$-wave
 ($s^{*}$) and $\phi_{k}=\eta_{k}=\cos(k_{x}a)-\cos(k_{y}a)$ for the 
 $d_{x^2-y^2}$-wave pairing.
 In general, one can consider a decomposition 
 $I\phi_{k}=g_{0}+g_{s}\gamma_{k}+g_{d}\eta_{k}$, 
 with suitable  coupling constants for different 
 symmetry channels.
Our analysis is based on  the BCS-Mean-Field Approximation (MFA) and the
 Kosterlitz-Thouless (KT) theory for 2D superfluid. 
 We do not consider here the direct bosonic hopping $J_{ij}$.
The superconducting state is characterized by  two order parameters: $x_{0}=
\frac{1}{N}\sum_{k}\phi_{k}<c^{\dagger}_{k\uparrow}c^{\dagger}_{-k\downarrow}>$
 and $\rho^{x}_{0}=\frac{1}{2N}\sum_{i}<b^{\dagger}_{i}+b_{i}>$, 
 which satisfy the set of  equations:
 \begin{eqnarray}
 x_{0}=
 -\frac{1}{N}\sum_{\bf k}\frac{I\phi_{k}^2\rho^{x}_{0}}{2E_{\bf k}}\tanh(\beta
 E_{\bf k}/2),\\
 \rho^{x}_{0}=-\frac{Ix_{0}}{2\Delta}\tanh(\beta\Delta),~~
 n=n_{c}+2n_{b},
  \end{eqnarray}
where the quasiparticle energy is given by $E_{k}=\sqrt{\bar\epsilon_{k}^{2}+
\Delta_{k}^{2}}$, $\bar\epsilon_{k}=\epsilon_{k}-\mu$, 
$\Delta_{\bf {k}}^2=I^2\phi_{k}^2(\rho_{0}^{x})^2$.
$\Delta=\sqrt{(\Delta_{0}-\mu)^2+I^2x_{0}^2}$.  
The $c$-electron  dispersion is  
$\epsilon_{k}=-2t(\cos (k_{x}a)+\cos (k_{y}a))-
4t_{2}\cos (k_{x}a)\cos (k_{y}a)-\epsilon_{b}$
with the next nearest neighbour (nnn) hopping parameter $t_{2}$,
$\epsilon_{b}=min\epsilon_{k}$.
It should be noted that the energy gap in a wide band  is due 
to Bose condensate ($<b>\neq 0$) 
and well defined Bogolyubov quasiparticles exist in the superconducting phase.
The superfluid density, derived within the BCS scheme, 
is given  by $\rho_{s}^{\alpha} =
\frac{1}{2N}\sum_{k}
\left\{
\left(\frac{\partial\epsilon_{k}}{\partial k_{\alpha}}\right)^{2}
\frac{\partial f(E_{k})}{\partial E_{k}} 
+\frac{1}{2}\frac{\partial^{2}\epsilon_{k}}{\partial k_{\alpha}^{2}}
\left[
1-\frac{\bar\epsilon_{k}}{E_{k}}
\tanh\left(\frac{\beta E_{k}}{2}\right)
\right]\right\}, $
where  $f(E_{k})$ is the Fermi-Dirac
distribution function and $\alpha=x, y, z$. 
In the local limit one has  $\lambda^{-2}=(16\pi e^2/\hbar^2c^2)\rho_{s}$, 
where $\lambda$ is the London penetration.
Finally, the effect of phase fluctuations on the critical temperatures is  
evaluated  within the KT theory, i.e., from the relation for the universal 
jump of the superfluid stiffness $\rho_{s}$ at $T_{c}$: 
$\rho_{s}^{-}(T_{c})=\frac{2}{\pi}k_{B}T_{c}$.\\ 
We have performed an extended analysis of the phase diagrams and 
superfluid properties of the model (1) for different pairing
symmetries \cite{unpublished}. 
The generic phase diagram (for  $d$-wave symmetry) plotted as a function 
of the position of  LP level $\Delta_{0}$ is shown in Fig.1. 
A sharp drop in the superfluid stiffness (and in the  KT transition 
temperature) occurs when the bosonic level reaches the bottom 
of the $c$-electron band and the system approaches the LP limit.  
In the opposite, BCS like limit, $T_{c}$ approaches
asymptotically the MF transition temperature, with  narrow fluctuation regime.
Between the KT and MFA temperatures the phase fluctuation effects are important.
In this regime a pseudogap in c-electron spectrum will develop and 
the normal state of LP and itinerant fermions  
can exhibit non-Fermi liquid properties \cite{jrmr96}.\\ 
With varying $n$ but for fixed  $\Delta_{0}$, it appears that 
the mechanism of induced superconductivity 
in the mixed regime of coexisting LP and electrons is not very
sensitive to the pairing symmetry, i.e. $n_{c}$ is nearly constant, 
but $n_{b}$ (LP level occupation) increases with total $n$.
The  chemical potential in the superconducting  phase is practically 
pinned around $\Delta_{0}$.
The superfluid density exhibits linear in T behavior 
(at low T) for $d_{x^2-y^2}$-wave pairing due to the existence of 
nodal quasiparticles. 
For the same  pairing symmetry, we have also found that the 
scaled stiffness $\rho_{s}(T)/\rho_{s}(0)$ vs $T/T_{c}$ 
shows only a weak dependence on the total density $n$. 
The d-wave pairing is favored for higher concentration of c electrons, while
the extended $s$-wave can be realized for lower $n_{c}$ (for the nn hopping).
The nnn hopping $t_{2}$ can substantially  enhance $T_{c}$ for $d$-wave 
symmetry.\\
Finally,  within the KT  scenario, 
the Uemura-type plots have been  obtained for $s^{*}$ and $d$--wave symmetry 
and an example is  shown in Fig.2. A linear scaling of $T_{c}$ (KT)  
with $\rho_{s}(0)$ is a result of separation of the energy scales 
for the pairing and phase coherence, which occurs in the presence  
of LP.
 
This work was supported in part by the Committee of Scientific Research (KBN) 
of Poland, Project No. 2 P03B 037 17.
\vfill

\eject

\vfill

\eject

\begin{figure}
\includegraphics{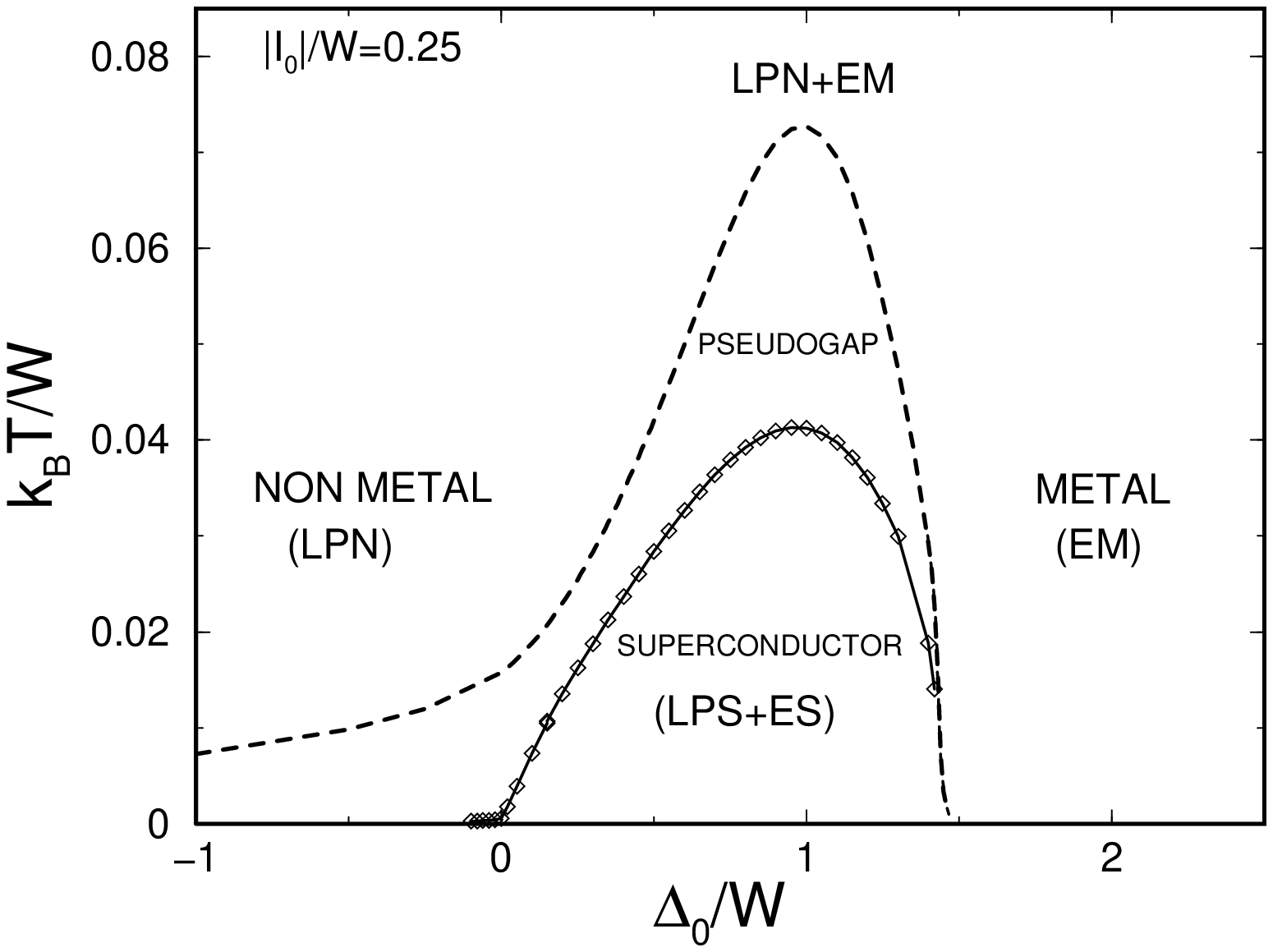}
\caption{Phase diagram of the induced pairing model for the 
$d_{x^2-y^2}$ -- wave  symmetry and   $n=1.5$.
 $I=-|I_{0}|$, $J_{ij}=0$. $W=4t$. The dashed line -- BCS-MFA transition
temperature,  the line with symbols -- KT transition temperature. 
LPN--nonmetallic phase of LP, EM--electronic metal, 
LPS+ES--superconducting state. }
\end{figure}

\begin{figure}[t]
\includegraphics{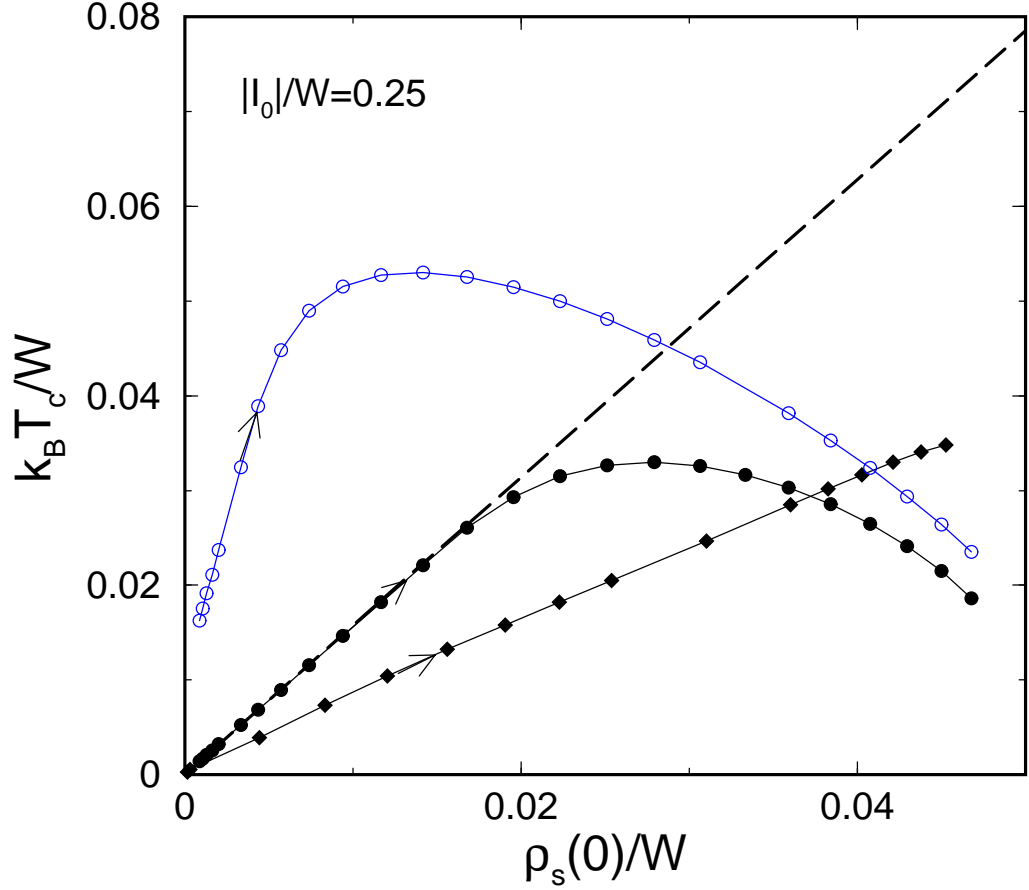}
\caption{Uemura type plots for fixed  density $n=1$. $t_{2}=0$. 
The parameter which drives
crossover from LP  to BCS limit is $\Delta_{0}$/W
and it increases from $-0.4$ to $0.85$ as 
shown by arrows. 
Circles correspond to the  $s^{*}$-wave pairing (empty-MFA, filled-KT), 
while  diamonds to the  $d$-wave pairing (KT). 
The dashed line $\pi\rho_{s}(0)/2W$ is an 
upper bound for phase ordering temperature.}
\end{figure}

\vfill
\end{document}